\documentclass[a4paper,11pt]{article}
\usepackage{pos}

    \usepackage{array}
    \usepackage{multirow}

\newcommand{\PreserveBackslash}[1]{\let\temp=\\#1\let\\=\temp}
\newcolumntype{C}[1]{>{\PreserveBackslash\centering}p{#1}}
\newcolumntype{R}[1]{>{\PreserveBackslash\raggedleft}p{#1}}
\newcolumntype{L}[1]{>{\PreserveBackslash\raggedright}p{#1}}

\title{Interpretation of LHCb Hidden-Charm Pentaquarks within the Compact Diquark Model}

\author[a]{Ahmed Ali}
\author[b]{Ishtiaq Ahmed}
\author[c]{M. Jamil Aslam} 
\author*[d]{Alexander Parkhomenko}
\author[b]{Abdur Rehman}

\affiliation[a]{Deutsches Elektronen-Synchrotron DESY, \\ 
  D-22607 Hamburg, Germany}

\affiliation[b]{National Centre for Physics, Quaid-i-Azam University Campus, \\ 
  Islamabad 45320, Pakistan} 

\affiliation[c]{Physics Department, Quaid-i-Azam University, \\  
  Islamabad 45320, Pakistan} 

\affiliation[d]{Department of Theoretical Physics, P.\,G.~Demidov Yaroslavl State University, \\ 
  Sovietskaya 14, 150003 Yaroslavl, Russia}

\emailAdd{ahmed.ali@desy.de}
\emailAdd{ishtiaqmusab@gmail.com}
\emailAdd{muhammadjamil.aslam@gmail.com}
\emailAdd{parkh@uniyar.ac.ru}
\emailAdd{Abdur.Rehman@fuw.edu.pl}

\renewcommand{\printHeadAuthors}{A.~Ali et al.}

\abstract{
The LHCb collaboration have recently updated their analysis of the resonant $J/\psi\, p$ mass 
spectrum in the decay $\Lambda_b^0 \to J/\psi\, p\, K^-$, making use of their combined Run~1 
and Run~2 data. In the updated analysis, three narrow states, $P_c (4312)^+$, $P_c (4440)^+$, 
and $P_c (4457)^+$, are observed. The spin-parity assignments of these states are not yet known. 
We interpret these narrow resonances as compact hidden-charm diquark-diquark-antiquark pentaquarks. 
Using an effective Hamiltonian, based on constituent quarks and diquarks, we calculate 
the pentaquark mass spectrum for the complete $SU (3)_F$ lowest $S$- and $P$-wave multiplets, 
taking into account dominant spin-spin, spin-orbit, orbital and tensor interactions. The resulting 
spectrum is very rich and we work out the quark flavor compositions, masses, and $J^P$ quantum numbers 
of the pentaquarks. However, heavy quark symmetry restricts the observable states in $\Lambda_b$-baryon,
as well as in the decays of the other weakly-decaying $b$-baryons, $\Xi_b$ and $\Omega_b$. 
In addition, some of the pentaquark states are estimated to lie below the $J/\psi\, p$ threshold 
in $\Lambda_b$-decays (and corresponding thresholds in $\Xi_b$- and $\Omega_b$-decays). They decay 
via $c \bar c$ annihilation into light hadrons or a dilepton pair, and are expected to be narrower 
than the $P_c$-states observed. We anticipate their discovery, as well as of the other pentaquark 
states present in the spectrum at the LHC, and in the long-term future at a Tera-$Z$ factory. 
}

\FullConference{%
  40th International Conference on High Energy physics - ICHEP2020\\
  July 28 - August 6, 2020\\
  Prague, Czech Republic (virtual meeting)
}

\begin{document}
\maketitle


Recently, the LHCb collaboration have presented an updated account of the resonant $J/\psi\, p$ 
mass spectrum in the decay $\Lambda_b^0 \to J/\psi\, p\, K^-$, based on the combined Run~1 and 
Run~2 data, adding up to 9~fb$^{-1}$~\cite{Aaij:2019vzc}. In this analysis, which supersedes 
their earlier findings from 2015~\cite{Aaij:2015tga}, nominal fits of the data have been performed 
with an incoherent sum of Breit-Wigner amplitudes, which have resulted in the observation of three 
peaks $P_c (4312)^+$, $P_c (4440)^+$, and $P_c (4457)^+$ with masses $4311.9 \pm 0.7^{+6.8}_{-0.6}$~MeV, 
$4440.3 \pm 1.3^{+4.1}_{-4.7}$~MeV, $4457.3 \pm 0.6^{+4.1}_{-1.7}$~MeV and corresponding decay widths 
$9.8 \pm 2.7^{+3.7}_{-4.5}$~MeV, $20.6 \pm 4.9^{+8.7}_{-10.1}$~MeV, $6.4 \pm 2.0^{+5.7}_{-1.9}$~MeV, 
respectively~\cite{Aaij:2019vzc}.
The state $P_c (4450)^+$ in the 2015 data~\cite{Aaij:2015tga}, is now replaced by two narrow states, 
$P_c (4440)^+$ and $P_c (4457)^+$. In addition, a third narrow peak, $P_c (4312)^+$, having the 
mass $M = (4311.9 \pm 0.7^{+6.8}_{-0.6})$~MeV, is also observed. The spin-parity,~$J^P$, assignments 
of the three narrow states, which are crucial to decipher the underlying dynamics of the pentaquark 
states, are not yet determined. The broad peak $P_c (4380)^+$ from the earlier data~\cite{Aaij:2015tga} 
is neither confirmed nor refuted, as the current LHCb analysis is not sensitive to broad resonances.

The constituents of the hidden-charm pentaquarks in the compact diquark model are~$[cq]_{\bar 3}$, 
$\bar c_{\bar 3}$, and~$[q^\prime q^{\prime\prime}]_{\bar 3}$, which make up a color singlet. 
However, it is still a three-body problem to solve and there are several dynamical possibilities 
to model their interconnections~\cite{Lebed:2015tna,Ali:2016dkf,Ali:2019npk,Ali:2019clg}. 
In~\cite{Ali:2019npk,Ali:2019clg}, we worked out the intuitive picture in which the heavier components 
form a nucleus and the lighter one is in an orbit around this nucleus, as it is energetically easier 
to excite light degrees of freedom. Spin-parity quantum numbers are fixed due to heavy-quark symmetry 
constraints. So, we keep the light diquark as emerging intact in $b$-baryon decays, and put it in the 
orbit for the $P$-wave states, with the heavier components, carrying a charm quark or charm antiquark, 
acting as a nucleus of (an almost) static color source. 

Mass estimates are worked out in an effective Hamiltonian approach, which apart from constituent 
quark and diquark masses, includes dominant spin-spin, spin-orbit, orbital, and tensor 
interactions~\cite{Ali:2019npk,Ali:2019clg}. Our mass predictions for the unflavored ($S = 0$), 
singly- ($S = -1$) and doubly-strange ($S = -2$) hidden-charm pentaquarks with a light spinless 
diquark ($S_{ld} = 0$) are collected in Table~\ref{tab:masses-predictions}.  
These states can be observed in weak decays of the bottom baryons. Masses of the states which are
candidates for the already 
observed  unflavored pentaquarks by the LHCb Collaboration are  put into boxes in the table. 
Note that the singly- and doubly-strange pentaquarks mentioned have a strange light diquark ($S = -1$), 
i.\,e. having the structures $(\bar c_{\bar 3} [c q]_{\bar 3} [s q^\prime]_{\bar 3})$ and 
$(\bar c_{\bar 3} [c s]_{\bar 3} [s q]_{\bar 3})$, where $q^{(\prime)}$ is $u$- or $d$-quark.  
Complete mass spectrum of the ground and orbitally excited hidden-charm pentaquarks can be found 
in~\cite{Ali:2019clg}. 

\begin{table}[tb] 
\caption{
Masses of the unflavored ($S = 0$), singly-strange ($S = -1$) and doubly-strange ($S = -2$) 
hidden-charm pentaquarks (in MeV) with a light diquark in the spin-0 state. 
}
\label{tab:masses-predictions}
\begin{center}
\begin{tabular}{|ccccc|} 
\hline
& $J^P$ & $S = 0$ & $S = -1$ &  $S = -1$ \\ \hline 
        & $1/2^-$ & $3830 \pm 34$            & $4112 \pm 32$  & $4243 \pm 32$ \\  
$L = 0$ &         & $4150 \pm 29$            & $4433 \pm 26$  & $4575 \pm 26$ \\  
        & $3/2^-$ & \framebox{$4240 \pm 29$} & $4523 \pm 26$  & $4644 \pm 26$ \\ \cline{1-4} 
        & $1/2^+$ & $4030 \pm 39$            & $4312 \pm 37$  & $4443 \pm 38$ \\  
        &         & $4351 \pm 35$            & $4633 \pm 33$  & $4775 \pm 33$ \\  
        &         & $4430 \pm 35$            & $4713 \pm 33$  & $4834 \pm 33$ \\  
$L = 1$ & $3/2^+$ & $4040 \pm 39$            & $4323 \pm 37$  & $4454 \pm 38$ \\  
        &         & $4361 \pm 35$            & $4643 \pm 33$  & $4785 \pm 33$ \\  
        &         & \framebox{$4440 \pm 35$} & $4723 \pm 33$  & $4844 \pm 33$ \\  
        & $5/2^+$ & \framebox{$4457 \pm 35$} & $4740 \pm 33$  & $4861 \pm 33$ \\ \hline 
\end{tabular}
\end{center} 
\end{table} 

We briefly comment on the recent evidence of a hidden-charm resonance, $P_{cs} (4459)$, 
with a strangeness $S = -1$. It was found in the $J/\psi\, \Lambda$ invariant mass spectrum 
in the $\Xi_b^- \to J/\psi\, \Lambda\, K^-$ decay with a statistical significance 
of~$3.1\sigma$~\cite{Wang:2020}. The Breit-Wigner mass and decay width of this state 
are measured to be $M_{P_{cs} (4459)} = 4458.8 \pm 2.9^{+4.7}_{-1.1}$~MeV and 
$\Gamma_{P_{cs} (4459)} = 17.3 \pm 6.5^{+8.0}_{-5.7}$~MeV but its spin and parity are not 
yet determined. Several assignments have been suggested for the $P_{cs} (4459)$-pentaquark 
and we also present a possible assignments based on the diquark-diquark-antiquark model 
of pentaquarks~\cite{Ali:2019clg}. As shown in Table~\ref{tab:masses-predictions}  
(see the second column with mass estimates), the mass of the $P_{cs} (4459)$-resonance 
is well correlated with the $J^P = 1/2^-$ state having the mass $M = 4433 \pm 26$~MeV.     


Experimental determination of the spins and parities of the LHCb pentaquarks will shed light 
on their underlying structures and allow us to discriminate among the suggested models for their description.

\acknowledgments{
A.\,P. acknowledges financial support by the Russian Foundation for Basic Research and National 
Natural Science Foundation of China according to the joint research project (No. 19-52-53041).
}

\bibliographystyle{JHEP}
\bibliography{Contrib-Parkhomenko-ID248} 



\end{document}